\documentclass[12pt]{article}
\usepackage{amsmath,amssymb,bm,graphicx}
\usepackage{color} 
\usepackage{hyperref}

\newcommand{\delslash}{\not \! \partial}

\begin{document}

\begin{center}
{\Large{\bf Note on a BCS analogy of Majorana neutrinos\footnote{The present note is dedicated to the memory of Lars Brink, who made important contributions to the International Journal of Modern Physics A as an Editor.}}}
\end{center}

\begin{flushright}
\end{flushright}

\vskip 0.5 truecm

\begin{center}
{\bf { Kazuo Fujikawa$^{1}$ and Anca Tureanu$^2$}}
\end{center}
\begin{center}
\vspace*{0.4cm} 
{\it {$^1$Interdisciplinary Theoretical and Mathematical Sciences Program (iTHEMS),\\
RIKEN, Wako 351-0198, Japan}
}\\
{\it {$^2$Department of Physics, University of Helsinki,\\
and Helsinki Institute of Physics,
\\P.O.Box 64, FIN-00014 Helsinki,
Finland}}
\end{center}
\makeatletter
\makeatother


\begin{abstract}

In this note, we discuss an analogy between the BCS theory and the seesaw model of neutrinos. We believe that the analogy indicates some fundamental aspects of Majorana neutrinos. A paper on the issue has been recently presented, and we would like to describe the background of the paper together with our personal views on the problem. In essence, the conventional construction of a Majorana neutrino from a chiral fermion is too simplified, and we argue that one would actually have to go through a Bogoliubov-type canonical  transformation to generate two Majorana fermions from an effective single Dirac fermion in the seesaw model.

\end{abstract}


\section {Introduction}

When one of us (KF) gave a seminar at Yanagida's group at IPMU, University of Tokyo, the other day, a young visitor from Berkeley commented after the seminar: `` It is interesing that the definition of   two Majorana fermions from a Dirac fermion has no problem. The definition of two chiral fermions from a Dirac fermion has no problem either. But the definition of a Majorana fermion from a single chiral fermion has many complications.'' This is precisely what we have been concerned with for several years. Recently, those considerations were summarized in a paper \cite{Fujikawa-Tureanu-2024}. We would like to briefly discuss the essence of the work together with our personal views about the problem in the following.

\section{ Possible charge conjugations of a chiral fermion}

It is well known that two Majorana fermions are defined in terms of a Dirac fermion $\psi(x)$ in the form:
\begin{eqnarray}
\psi_{M_{1}}&=&\frac{1}{\sqrt{2}}\{\psi(x)+\psi^{C}(x)\}\nonumber\\
\psi_{M_{2}}&=&\frac{1}{\sqrt{2}}\{\psi(x)-\psi^{C}(x)\},
\end{eqnarray}
where $\psi^{C}(x)$ is the charge conjugate of $\psi(x)$,
\begin{eqnarray}
\psi^{C}(x)=C\overline{\psi}^{T}(x),
\end{eqnarray}
with $C=i\gamma^{2}\gamma^{0}$.  Our notational conventions generally follow those in \cite{Bjorken}. The charge conjugations of Majorana fermions are 
\begin{eqnarray}
C: \psi^{C}_{M_{1}}(x)=\psi_{M_{1}}(x), \ \ \psi^{C}_{M_{2}}(x)=-\psi_{M_{2}}(x),
\end{eqnarray}
respectively.

It is also well known that a Dirac fermion is decomposed into a pair of chiral fermions,
\begin{eqnarray}
\psi(x)= \psi_{L}(x)+ \psi_{R}(x),
\end{eqnarray}
where $\psi_{R,L}(x)=(\frac{1\pm \gamma_{5}}{2})\psi(x)$. 

It is then interesting how to define the charge conjugation of chiral fermions. It is common to define the charge conjugation of chiral fermions by 
\begin{eqnarray}
C: \psi_{L}^{C}(x)=C\overline{\psi_{R}}^{T}(x), \ \ \psi_{R}^{C}(x)=C\overline{\psi_{L}}^{T}(x),
\end{eqnarray}
which is called C-conjugation in the following, whereas we call the conjugation defined by 
\begin{eqnarray}
\tilde{C}: \psi_{L}^{\tilde{C}}(x)=C\overline{\psi_{L}}^{T}(x), \ \ \psi_{R}^{\tilde{C}}(x)=C\overline{\psi_{R}}^{T}(x)
\end{eqnarray}
as pseudo-C symmetry.  If one looks at the Dirac field
$\psi(x)=\psi_{L}(x)+ \psi_{R}(x)$, these definitions give in a naive operator sense
\begin{eqnarray}
\psi(x)=\psi_{L}(x)+ \psi_{R}(x) \rightarrow \psi_{L}^{C}(x) +\psi_{R}^{C}(x)=(\psi_{R}(x)+\psi_{L}(x))^{C}=\psi(x)^{C}
\end{eqnarray}
and 
\begin{eqnarray}
\psi(x)=\psi_{L}(x)+ \psi_{R}(x) \rightarrow \psi_{L}^{\tilde{C}}(x) +\psi_{R}^{\tilde{C}}(x)=(\psi_{L}(x)+\psi_{R}(x))^{C}=\psi(x)^{C}
\end{eqnarray}
respectively, namely, either definition of charge conjugation is consistent with the basic definition of the charge conjugation of a Dirac fermion extracted from Quantum Electrodynamics.

The major difference of these two possible charge conjugations appears in their chirality properties. Under the conventional charge conjugation of a chiral fermion, denoted by an operator C,  we have 
\begin{eqnarray}
\psi_{L,R}(x)\rightarrow \psi^{C}_{L,R}=C\overline{\psi_{R,L}}^{T}(x)=
\{C\overline{\psi}^{T}\}_{L,R}(x),
\end{eqnarray}
namely, the overall chirality is preserved,  while we have 
\begin{eqnarray}
\psi_{L,R}(x)\rightarrow \psi^{\tilde{C}}_{L,R}=C\overline{\psi_{L,R}}^{T}(x)=
\{C\overline{\psi}^{T}\}_{R,L}(x)
\end{eqnarray}
namely, the overall chirality is interchanged under the pseudo-C 
 symmetry denoted by the operator $\tilde{C}$.

\section{Chirality changing transformation is not defined in Lagrangian field theory}
 
The notion of chirality may appear to be a local notion and thus one might guess that a chirality changing transformation is  local. But actually it is not realized in a local field theory \cite{Fujikawa-Tureanu}. 
An explicit physical example is a  {\em pseudo-C transformation} of a left-handed neutrino $N_{L}(x)$, which is denoted by $\tilde{C}$,
\begin{eqnarray}\label{2}
N_{L}(x)\rightarrow (N_{L})^{\tilde{C}}=C\overline{N_{L}}^{T}.
\end{eqnarray}
This symmetry satisfies 
$[(N_{L})^{\tilde{C}}]^{\tilde{C}}=N_{L}$.  Note that $(N_{L})^{\tilde{C}}=C\overline{N_{L}}^{T}$ is overall right-handed for a left-handed $N_{L}$.

The definition of a Majorana neutrino 
 $N=N_{L}+N_{R}$ with $N_{R}=C\overline{N_{L}}^{T}$ and $C=i\gamma^{2}\gamma^{0}$ is commonly used. This construction would appear to be natural (superficially) if one recalls  
\begin{eqnarray}\label{1}
N=N_{L}+N_{R}=N_{L}+C\overline{N_{L}}^{T}\equiv N_{L}+(N_{L})^{\tilde{C}}.
\end{eqnarray}
This symmetry is required to be satisfied always,  to ensure that $N(x)$ is a Majorana fermion by this definition. This pseudo-C transformation is also often regarded as the true C transformation in the literature \cite{Bilenky}.

But this construction leads to a contradiction if one defines an action in terms of \eqref{1},
\begin{eqnarray}\label{Vanishing Majorana}
\int d^{4}x {\cal L}&=&\int d^{4}x\frac{1}{2} \{\overline{N}(i\delslash -M)N\}\nonumber\\
&=&\int d^{4}x\frac{1}{2} \{\overline{N_{L}}i\delslash N_{L}
+\overline{C\overline{N_{L}}^{T}}i\delslash C\overline{N_{L}}^{T} -\overline{C\overline{N_{L}}^{T}}M N_{L} -\overline{N_{L}}M C\overline{N_{L}}^{T}\}\nonumber\\
&=&\int d^{4}x \frac{1}{2}\{\overline{N_{L}}i\delslash \frac{1-\gamma_{5}}{2} N_{L} +\overline{C\overline{N_{L}}^{T}}i\delslash \frac{1+\gamma_{5}}{2}C\overline{N_{L}}^{T}
\nonumber\\
&&\hspace{1.5cm} -\overline{C\overline{N_{L}}^{T}}M\frac{1-\gamma_{5}}{2} N_{L} -\overline{N_{L}}M \frac{1+\gamma_{5}}{2}C\overline{N_{L}}^{T}\}.
\end{eqnarray}
This action is supposed to be a Majorana-type and invariant under the charge conjugation. But this action completely  {\em vanishes} if one should use  
$N_{L}\rightarrow C\overline{N_{L}}^{T}$ and $C\overline{N_{L}}^{T}\rightarrow N_{L}$ in \eqref{2}, namely, if one should apply the chirality changing pseudo-C transformation law. 

This property generally arises from the chirality changing transformation law, since one can write 
\begin{eqnarray}
N(x) = N_{L}(x)+ C\overline {N_{L}}^{T}(x) = \frac{(1-\gamma_{5})}{2}N_{L}(x)+ \frac{(1+\gamma_{5})}{2}C\overline {N_{L}}^{T}(x)
 \end{eqnarray}
inside the local Lagrangian.  We thus have 
\begin{eqnarray}
N^{\tilde{C}}(x) &=& \frac{(1-\gamma_{5})}{2}{N_{L}}^{\tilde{C}}(x)+ \frac{(1+\gamma_{5})}{2}[C\overline {N_{L}}^{T}]^{\tilde{C}}(x)\nonumber\\
&=&\frac{(1-\gamma_{5})}{2}C\overline {N_{L}}^{T}(x)+ \frac{(1+\gamma_{5})}{2}N_{L}(x)=0.
 \end{eqnarray} 
 We emphasize that the field $N(x)$ by itself  is well-defined by CP symmetry but it vanishes always when the pseudo-C  transformation law \eqref{2} is enforced, namely, if one wants to identify $N(x)$ as a Majorana fermion in the sense of  pseudo-C symmetry\footnote{This reasoning is related to the Majorana neutrino as a Bogoliubov quasiparticle \cite{Fujikawa-Tureanu2}. If one should assume a local operator with ${\cal \tilde{C}}N_{L}(x){\cal \tilde{C}}^{\dagger}=C\overline {N_{L}}^{T}(x)$, then one would have ${\cal \tilde{C}}N_{L}(x){\cal \tilde{C}}^{\dagger}=(\frac{1-\gamma_{5}}{2}){\cal \tilde{C}}N_{L}(x){\cal \tilde{C}}^{\dagger}=(\frac{1-\gamma_{5}}{2})C\overline {N_{L}}^{T}(x)=0$. The definition of a well-defined charge for the pseudo-C is subtle since the Lagrangian is not well defined, and we do not use it in the present note.}.

  One may compare the two different charge conjugation symmetries with explicit projection operators in a local Lagrangian; the conventional C gives 
 \begin{eqnarray}   
 \psi^{C}&=&\frac{1-\gamma_{5}}{2}\psi^{C}_{L}  +\frac{1+\gamma_{5}}{2}\psi^{C}_{R}=\frac{1-\gamma_{5}}{2}C\overline{\psi_{R}}^{T} +\frac{1+\gamma_{5}}{2}C\overline{\psi_{L}}^{T}\nonumber\\
 &=&C\overline{(\frac{1+\gamma_{5}}{2})\psi_{R}+ (\frac{1-\gamma_{5}}{2})\psi_{L}}^{T}=C\overline{\psi}^{T},
  \end{eqnarray}
   while the pseudo-C symmetry  
   \begin{eqnarray}
    \psi^{\tilde{C}}=(\frac{1-\gamma_{5}}{2})\psi^{\tilde{C}}_{L}  +(\frac{1+\gamma_{5}}{2})\psi^{\tilde{C}}_{R}=(\frac{1-\gamma_{5}}{2})C\overline{\psi_{L}}^{T} +(\frac{1+\gamma_{5}}{2})C\overline{\psi_{R}}^{T}=0.
\end{eqnarray}
In Lagrangian field theory, the conventional C is well-defined but the pseudo-C is not consistently defined.

 \section{Seesaw model and the Bogoliubov-type transformation} 
 
It is standard to use the seesaw model \cite{Minkowski,Yanagida,Mohapatra, Fukugita} to generate  Majorana fermions in an extended Standard Model. The seesaw model is defined by 
\begin{eqnarray}\label{Seesaw}
{\cal L} &=& \overline{\nu_{L}}(x)i\delslash\nu_{L}(x)
+\overline{\nu_{R}}(x)i\delslash\nu_{R}(x)\nonumber\\
&&- \{ \overline{\nu_{L}}m_{D}\nu_{R}(x) 
+(1/2)\nu^{T}_{R}(x)Cm_{R}\nu_{R}(x)+h.c.\},
\end{eqnarray}
where $m_{D}$ is a $3\times 3$ complex Dirac-type  mass matrix, and $m_{R}$ is a  $3\times 3$ complex symmetric matrix. The anti-symmetry of the matrix $C=i\gamma^{2}\gamma^{0}$
 and Fermi statistics imply that $m_{R}$ is symmetric. This is the Lagrangian of neutrinos with Dirac and Majorana mass terms, and it represents the classical seesaw Lagrangian of type I; namely, starting with the Dirac-type neutrinos, the very large masses $m_{R}$ are added to gauge-singlets $\nu_{R}$ and thus intuitively making $\nu_{R}$ very massive, in a manner consistent with the gauge structure of the SM.  
 
We start with the Lagrangian \eqref{Seesaw}, which is CP invariant but with no obvious C or (left-right) P symmetry, and write the mass term as 
\begin{eqnarray}\label{mass term}
(-2){\cal L}_{mass}=
\left(\begin{array}{cc}
            \overline{\nu_{R}}&\overline{\nu_{R}^{C}}
            \end{array}\right)
\left(\begin{array}{cc}
            m_{R}& m_{D}\\
            m_{D}^{T}&0
            \end{array}\right)
            \left(\begin{array}{c}
            \nu_{L}^{C}\\
            \nu_{L}
            \end{array}\right) +h.c.,
\end{eqnarray}
where 
\begin{eqnarray}\label{notational convention1}
\nu_{L}^{C}\equiv C\overline{\nu_{R}}^T, \ \ \ \nu_{R}^{C}\equiv C\overline{\nu_{L}}^T.  
\end{eqnarray}
Since the mass matrix is complex symmetric, we can diagonalize it precisely
by a $6 \times 6$ unitary Autonne--Takagi \cite{Autonne--Takagi} factorization  as
\begin{eqnarray}\label{orthogonal}
            U^{T}
            \left(\begin{array}{cc}
            m_{R}& m_{D}\\
            m_{D}^{T}& 0
            \end{array}\right)
            U
            =\left(\begin{array}{cc}
            M_{1}&0\\
            0&-M_{2}
            \end{array}\right)    ,        
\end{eqnarray}
where  $M_{1}$ and $M_{2}$ are $3\times 3$ real diagonal matrices.
Hence we can write the seesaw Lagrangian in the form 
\begin{eqnarray}\label{exact-solution}
{\cal L}
 &=&(1/2)\{\overline{\psi_{+}}[i\delslash  -M_{1}]\psi_{+} +\overline{\psi_{-}}[i\delslash  - M_{2}]\psi_{-}\} ,
\end{eqnarray}
where
\begin{eqnarray}\label{pseudoMajorana}
\psi_{+}(x)&=& \tilde{\nu}_{R}+ \tilde{\nu}_{L}^{C}=\tilde{\nu}_{R}+ C\overline{\tilde{\nu}_{R}}^{T},\nonumber\\
\psi_{-}(x)&=&\tilde{\nu}_{L}- \tilde{\nu}_{R}^{C}= \tilde{\nu}_{L}- C\overline{\tilde{\nu}_{L}}^{T}.
\end{eqnarray}
Here we used 
\begin{eqnarray} \label{variable-change}          
            &&\left(\begin{array}{c}
            \nu_{L}^{C}\\
            \nu_{L}
            \end{array}\right)
            = U \left(\begin{array}{c}
            \tilde{\nu}_{L}^{C}\\
            \tilde{\nu}_{L}
            \end{array}\right)
           ,\ \ \ \ 
            \left(\begin{array}{c}
            \nu_{R}\\
            \nu_{R}^{C}
            \end{array}\right)
            = U^{\star} 
            \left(\begin{array}{c}
            \tilde{\nu}_{R}\\
            \tilde{\nu}_{R}^{C}
            \end{array}\right),         
\end{eqnarray}
and mass eigenvalues $M_{1}\gg M_{2}$.

The ``Majorana'' neutrinos \eqref{pseudoMajorana} are defined by the pseudo-C symmetry which is not consistently defined as we have explained; the Majorana neutrinos \eqref{pseudoMajorana} are well-defined by CP symmetry but not by the pseudo-C symmetry. Our strategy is then to apply a generalized Pauli--G\"{u}rsey transformation \cite{Pauli, KF-PG}, which is an analogue of the Bogoliubov transformation \cite{Bogoliubov} in BCS theory, to the seesaw Lagrangian \eqref{exact-solution}.   
We thus consider a further $6 \times 6$ real generalized Pauli--G\"{u}rsey transformation $O$, which has generally  the same form as the Autonne--Takagi factorization, but it is orthogonal in the present specific case and thus it preserves CP:
\begin{eqnarray} \label{Pauli--Gursey2}          
            &&\left(\begin{array}{c}
            \tilde{\nu}_{L}^{C}\\
            \tilde{\nu}_{L}
            \end{array}\right)
            = O \left(\begin{array}{c}
            N_{L}^{C}\\
            N_{L}
            \end{array}\right)
           ,\ \ \ \ 
             \left(\begin{array}{c}
            \tilde{\nu}_{R}\\
            \tilde{\nu}_{R}^{C}
            \end{array}\right)
            = O 
            \left(\begin{array}{c}
            N_{R}\\
            N_{R}^{C}
            \end{array}\right),         
\end{eqnarray}
with a specific $6 \times 6$ orthogonal transformation 
\begin{eqnarray}\label{Pauli-Gursey}
O=\frac{1}{\sqrt{2}} \left(\begin{array}{cc}
            1&1\\
            -1&1
            \end{array}\right).
\end{eqnarray}
 The seesaw Lagrangian is then transformed to 
\begin{eqnarray}\label{seeaw Lagrangian3}
{\cal L}
&=&(1/2)\{\overline{\Psi_{+}}[i\delslash  -M_{1}]\Psi_{+} +\overline{\Psi_{-}}[i\delslash  - M_{2}]\Psi_{-}\},
\end{eqnarray}
where 
\begin{eqnarray}
\Psi_{+}=\frac{1}{\sqrt{2}} (N_{R} +N_{R}^{C})+\frac{1}{\sqrt{2}}(N_{L} +N_{L}^{C})=\frac{1}{\sqrt{2}} (N +N^{C}),\nonumber\\
\Psi_{-}=\frac{1}{\sqrt{2}}(N_{L} -N_{L}^{C})-\frac{1}{\sqrt{2}}(N_{R}^{C}-N_{R})=\frac{1}{\sqrt{2}}(N -N^{C}).
\end{eqnarray}
which are invariant under the standard C, P and CP defined by 
\begin{eqnarray}
&&C:\ \ N(x)\leftrightarrow N^{C}(x)=C\overline{N}^{T}(x)\nonumber\\
&&P:\ \ N(t,\vec{x})\rightarrow i\gamma^{0}N(t,-\vec{x}),\ \ N^{C}(t,\vec{x})\rightarrow i\gamma^{0}N^{C}(t,-\vec{x}),\nonumber\\
&&CP:\ \ N(x)\rightarrow i\gamma^{0}N^{C}(t,-\vec{x}),\ \ N^{C}(x)\rightarrow i\gamma^{0}N(t,-\vec{x}), 
\end{eqnarray}
namely, $N(t,\vec{x})$ are Dirac-type variables. The phase of parity  $i\gamma^{0}$ is chosen for the convenience to deal with a Majorana fermion \cite{Weinberg0}.

In the generalized Pauli--G\"{u}rsey canonical transformation, we have
\begin{eqnarray}\label{Pauli-Gursey2}
\psi_{+}&=&\tilde{\nu}_{R}+C\overline{\tilde{\nu}_{R}}^{T}\Rightarrow \frac{1}{\sqrt{2}} (N_{R} +N_{R}^{C})+\frac{1}{\sqrt{2}}(N_{L} +N_{L}^{C})=\Psi_{+}\nonumber\\
\psi_{-}&=&\tilde{\nu}_{L}-C\overline{\tilde{\nu}_{L}}^{T}\Rightarrow 
\frac{1}{\sqrt{2}}(N_{L} -N_{L}^{C})-\frac{1}{\sqrt{2}}(N_{R}^{C}-N_{R})=\Psi_{-}.
\end{eqnarray}
Before the canonical transformation, $\tilde{\nu}_{R}$ and $C\overline{\tilde{\nu}_{R}}^{T}$, for example, are related by the pseudo-C with opposite chiralities, while these two terms are mapped to $\frac{1}{\sqrt{2}} (N_{R} +N_{R}^{C})$ and $\frac{1}{\sqrt{2}}(N_{L} +N_{L}^{C})$ with Dirac-type variables $N$, respectively, which are separately invariant under the conventional C with opposite chiralities; the CP symmetry is common for both sides in \eqref{Pauli-Gursey2}
but C and P are different.

Our view is that this Bogoliubov-type canonical transformation is inevitable in the consistent treatment of the seesaw Lagrangian \eqref{Seesaw} in an extension of SM.

 \section{ Weak interactions}
Let us recall the charged current weak interaction vertex in the SM,
 \begin{eqnarray}
 {\cal L}_{W}=\frac{g}{\sqrt{2}}\overline{l_{L}}(x)\gamma^{\mu}W_{\mu}^{-}(x)\frac{(1-\gamma_{5})}{2}\nu_{L}(x) + h.c.,
 \end{eqnarray}
 where $l_{L}(x)$ and $\nu_{L}(x)$ are the charged lepton triplet and the neutrino  triplet (but belonging to the gauged  $SU_{L}(2)\times U(1)$) in \eqref{Seesaw}, respectively. 
 This interaction Lagrangian after diagonalizing the charged lepton mass and the neutrino sector by \eqref{variable-change} is given by 
 \begin{eqnarray}\label{weak interaction}
 {\cal L}_{W}
 &=&\frac{g}{\sqrt{2}}\overline{l_{L}}^{\alpha}(x)\gamma^{\mu}W_{\mu}^{-}(x)[U^{\alpha k}\tilde{\nu}^{k}_{L}(x)+(U^{\prime})^{\alpha k}(\tilde{\nu}_{L}^{C})^{k}(x)] + h.c..
 \end{eqnarray}
The $3\times 3$ mixing matrices $U^{\alpha k}$ and $(U^{\prime})^{\alpha k} $ contain the products of the charged leptonic mixing matrix and the parts of the $3\times 3$ submatrices ($U_{22}$ and $U_{21}$, respectively) of $U$ in the first relation of \eqref{variable-change}. The mixing matrix $U^{\alpha k}$ may be identified with the conventional PMNS mixing matrix.  One may  rewrite \eqref{weak interaction} in the form
 \begin{eqnarray}\label{weak interaction2}
 {\cal L}_{W}
 &=&\frac{g}{\sqrt{2}}\overline{l_{L}}^{\alpha}(x)\gamma^{\mu}W_{\mu}^{-}(x)U^{\alpha k}\frac{(1-\gamma_{5})}{2}(\tilde{\nu}_{L}(x)-C\overline{\tilde{\nu}_{L}}^{T})^{k} \nonumber\\ &+& \frac{g}{\sqrt{2}}\overline{l_{L}}^{\alpha}(x)\gamma^{\mu}W_{\mu}^{-}(x)[(U^{\prime})^{\alpha k}(\tilde{\nu}_{L}^{C})^{k}(x)] +h.c.
 \end{eqnarray}
where we added the {\em right-handed} component $-C\overline{\tilde{\nu}_{L}}^{T}$, which is cancelled by $\frac{(1-\gamma_{5})}{2}$ in front, 
to form a Majorana fermion in \eqref{pseudoMajorana},
\begin{eqnarray}\label{free Majorana2}
\psi_{-}(x)=\tilde{\nu}_{L}(x)-C\overline{\tilde{\nu}_{L}}^{T}.
\end{eqnarray}
By assuming that the {\em left-handed} $\tilde{\nu}_{L}^{C}$ is very massive and thus may be ignored in the present discussions, one may  
identify \eqref{free Majorana2} with a Majorana neutrino in the double beta decay. To ensure being a Majorana neutrino, the pseudo-C symmetry in \eqref{1} needs to be imposed, but then the free part of the Majorana neutrino action and thus the free propagator vanish as in \eqref{Vanishing Majorana}. This means that no neutrinoless double beta decay would take place in the seesaw model \cite{Fujikawa}, at least if one enforces the pseudo-C symmetry to identify a Majorana neutrino\footnote{This argument applies also to the model of Weinberg \cite{Weinberg},
although the neutrinoless double beta decay takes place in Weinberg's model due to the violation of the lepton number in a CP invariant theory. But the ``Majorana'' neutrino appearing there  is not consistetly defined by the pseudo-C symmetry. Note that the Majorana neutrino and the lepton number non-conservation are not identical in the present context.}.

Another characteristic property of the ``Majorana'' neutrinos is that there appear extra CP violating phases \cite{Bilenky, Schechter2}. These extra CP phases also vanish since the action vanishes if one 
employs the first term in \eqref{weak interaction2} and imposes the Majorana nature by the pseudo-C symmetry.

In contrast, after the Bogoliubov-type tansformation \eqref{Pauli-Gursey2}, the weak interaction vertex is written in terms of conventional Majorana neutrinos as \cite{Fujikawa-Tureanu-2024}
 \begin{eqnarray}\label{weak interaction3}
 {\cal L}_{W}
 &=&\frac{g}{\sqrt{2}}\overline{l_{L}}^{\alpha}(x)\gamma^{\mu}W_{\mu}^{-}(x)[U^{\alpha k}{\Psi_{-}^{k}}_{L}(x)+(U^{\prime})^{\alpha k}{\Psi_{+}^{k}}_{L}(x)] + h.c..
 \end{eqnarray}  
and we have both the neutrinoless double beta decay and the extra CP phases in a natural manner. Note that one may ignore the term with $\Psi_{+}(x)$ which is assumed to be very heavy.

The absence of a Majorana-Weyl fermion in D=4 is suggestive.

\section{Two classes of Majorana fermions in the seesaw model}

To argue that the generalized Pauli--G\"{u}rsey canonical transformation (or Bogoliubov-type transformation) is very essential in the seesaw model, it is important to see the appearance of two classes of Majorana fermions in a general formulation. For this purpose, it is 
instructive  to look at the problem from a reversed direction,  using the notational conventions with  $\gamma_{5}$ diagonal,
\begin{eqnarray}\label{conventional metric conventions}
\vec{\gamma}=\left(\begin{array}{cc}
            0&-i\vec{\sigma}\\
            i\vec{\sigma}&0
            \end{array}\right),
\gamma_{4}=\left(\begin{array}{cc}
            0&1\\
            1&0
            \end{array}\right),  
\gamma_{5}=\left(\begin{array}{cc}
            1&0\\
            0&-1
            \end{array}\right),  
C=\left(\begin{array}{cc}
            -\sigma_{2}&0\\
            0&\sigma_{2}
            \end{array}\right).
\end{eqnarray}            
The 4-component Dirac field in a theory of  single species is then parameterized by 
\begin{eqnarray}\label{4-component Dirac}
&&\psi(x) =\left(\begin{array}{c}
            \chi\\
            \sigma_{2}\phi^{\star}
            \end{array}\right) , \ \  
\psi^{C}(x) =C\overline{\psi}^{T}(x)=\left(\begin{array}{c}
            \phi\\
            \sigma_{2}\chi^{\star}
            \end{array}\right).     
 \end{eqnarray}  
 The C transformation implies $\chi \leftrightarrow \phi$, namely, the chiral picture is
 \begin{eqnarray}
 C:\   \psi_{R}=\left(\begin{array}{c}
            \chi\\
            0
            \end{array}\right) \rightarrow                  
 C\overline{\psi_{L}}^{T}=\left(\begin{array}{c}
            \phi\\
            0
            \end{array}\right), \
 \psi_{L}=\left(\begin{array}{c}
            0\\
            \sigma_{2}\phi^{\star}
            \end{array}\right) \rightarrow                  
 C\overline{\psi_{R}}^{T}=\left(\begin{array}{c}
            0\\
            \sigma_{2}\chi^{\star}
            \end{array}\right). 
\end{eqnarray}            
The conventional Majorana fermions are defined by a rotation of $\pi/4$,
\begin{eqnarray}\label{ordinary transformation2}
\chi=\frac{1}{\sqrt{2}}(\rho_{1}+\rho_{2}),\ \
\phi=\frac{1}{\sqrt{2}}(\rho_{1}-\rho_{2}),
\end{eqnarray}
and their C transformation properties are  $\rho_{1}\rightarrow \rho_{1}$ and $\rho_{2}\rightarrow -\rho_{2}$. When one defines
\begin{eqnarray}\label{conventional Majorana}
&&\psi(x)=\frac{1}{\sqrt{2}}(\psi_{M_{1}}+\psi_{M_{2}}),\ \ 
\psi_{M_{1}}=\left(\begin{array}{c}
            \rho_{1}\\
            \sigma_{2}\rho_{1}^{\star}
            \end{array}\right), \ \ 
\psi_{M_{2}}=\left(\begin{array}{c}
            \rho_{2}\\
            -\sigma_{2}\rho_{2}^{\star}
            \end{array}\right),\nonumber\\
 && \psi^{C}(x)=\frac{1}{\sqrt{2}}(\psi_{M_{1}}-\psi_{M_{2}}),          
\end{eqnarray}                     
they satisfy the properties of the conventional Majorana fermions,
\begin{eqnarray}
&&\psi_{M_{1}}=C\overline{\psi_{M_{1}}}^{T}, \ \  \psi_{M_{2}}=-C\overline{\psi_{M_{2}}}^{T},
\end{eqnarray}
consistent with the C-transformation properties of $\rho_{1}$ and $\rho_{2}$. 

In this notation, another class of ``Majorana'' fermions are defined by 
\begin{eqnarray}\label{extra Majorana}
\psi_{+}(x)=\psi_{R}(x)+ C\overline{\psi_{R}(x)}^T=\left(\begin{array}{c}
            \chi\\
            \sigma_{2}\chi^{\star}
            \end{array}\right),\ 
\psi_{-}(x)=\psi_{L}(x)- C\overline{\psi_{L}(x)}^T=\left(\begin{array}{c}
            -\phi\\
            \sigma_{2}\phi^{\star}
            \end{array}\right),
\end{eqnarray}
which satisfy formally the Majorana conditions $\psi_{+}=C\overline{\psi_{+}}^{T}$ and $\psi_{-}=-C\overline{\psi_{-}}^{T}$, but the pseudo-C transformation $\psi_{L}\rightarrow C\overline{\psi_{L}}^{T}$, for example,  is different from what we expect for $\phi$; for this reason, besides the pseudo-C symmetry being not defined in Lagrangian field theory, these fermions are usually discarded. 

We have argued in \cite{Fujikawa-Tureanu-2024} that the exact diagonalization of the seesaw model leads to the class of fermions in \eqref{extra Majorana}. We then applied the generalized Pauli--G\"{u}rsey canonical transformation (or Bogoliubov-type transformation) in  \eqref{Pauli-Gursey2} written in the present context
\begin{eqnarray}
\left(\begin{array}{c}
            {\psi_{+}}_{R}\\
            {\psi_{+}}_{L}
            \end{array}\right)=\left(\begin{array}{c}
            \tilde{\nu}_{R}\\
            \tilde{\nu}_{L}^{C}
            \end{array}\right)=\left(\begin{array}{c}
            \chi\\
            \sigma_{2}\chi^{\star}
            \end{array}\right)
            &\Rightarrow&
            \left(\begin{array}{c}
           \rho_{1}\\
            \sigma_{2}\rho_{1}^{\star}
            \end{array}\right)
            =\left(\begin{array}{c}
            {\Psi_{+}}_{R}\\
            {\Psi_{+}}_{L}
            \end{array}\right),\nonumber\\
\left(\begin{array}{c}
            {-\psi_{-}}_{R}\\
            {\psi_{-}}_{L}
            \end{array}\right)=\left(\begin{array}{c}
            \tilde{\nu}_{R}^{C}\\
            \tilde{\nu}_{L}
            \end{array}\right)=\left(\begin{array}{c}
            \phi\\
            \sigma_{2}\phi^{\star}
            \end{array}\right)
            &\Rightarrow&         
            \left(\begin{array}{c}
            -\rho_{2}\\
            -\sigma_{2}\rho_{2}^{\star}
            \end{array}\right)
            =\left(\begin{array}{c}
            -{\Psi_{-}}_{R}\\
            {\Psi_{-}}_{L}
            \end{array}\right), 
\end{eqnarray}
where the double arrow indicates a generalized Pauli--G\"{u}rsey (or Bogoliubov-type) canonical transformation to generate  the conventional
Majorana neutrinos in \eqref{conventional Majorana}.

\section{Discussion}

It has been shown in \cite{Fujikawa-Tureanu-2024}   that the Majorana neutrinos in the seesaw model are most naturally defind using the Bogoliubov-type transformation. Our argument is based on the fact that the exact diagonalization of the seesaw model leads to neutrinos defined by the pseudo-C symmetry. The action for the Majorana neutrinos based on the pseudo-C symmetry is not defined in Lagrangian field theory. In essence, the Majorana neutrinos consistent in the Lagrangian field theory are limited to those conventional Majorana fermions defined using the Dirac-type fermions. We think that  the appearance of the analogy of BCS theory in the seesaw model is interesting. 

As for the related past work, we mention the formulation by Schechter and Valle on the massive Majorana neutrinos \cite{Schechter}. They formulated the Majorana neutrinos directly in the classical seesaw-like model without going through the pseudo-C symmetry. Our view is that the formulation using the Bogoliubov-type transformation is more natural since one can in principle define two classes of Majorana neutrinos in a vector-like theory such as the seesaw model \cite{Fujikawa-Tureanu-2024}.
 
The experimental determination if the neutrino is a Dirac fermion or a Majorana fermion is awaited.

\end{document}